\documentclass[12pt]{article}

\usepackage{epsf}
\usepackage{amsmath}
\usepackage{amsfonts}
\usepackage[mathscr]{euscript}

\newcommand{\np}[3]{Nucl. Phys. {\bf B#1}, #3 (19#2)}
\newcommand{\pl}[3]{Phys. Lett. {\bf B#1}, #3 (19#2)}
\newcommand{\pr}[3]{Phys. Rev. {\bf D#1}, #3 (19#2)}

\newcommand{\vj}[4]{#1~{\bf #2}, #4 (19#3)}

\def\be{\begin{equation}}
\def\ee{\end{equation}}
\def\bea{\begin{eqnarray}}
\def\eea{\end{eqnarray}}

\newcommand{\nn}{\nonumber\\}

\def\etal{{\it et al.\/}}

\def\ie{{\it i.e.\/}}

\newcommand{\la}{\langle}
\newcommand{\ra}{\rangle}
\newcommand{\Tr}{\mbox{Tr}}
\newcommand{\Str}{\mbox{Str}}
\newcommand{\N}{$\cal N$}

\begin{document}

\pagestyle{empty} 
{\hfill \parbox{6cm}{\begin{center} 
        MIT-CTP-2933  \\ 
        hep-th/9912250 \\
        December 1999                   
\end{center}}} 
              
\vspace*{1cm}                               
\begin{center} 
\large{\bf Non-renormalization of next-to-extremal correlators \\ }
\large{\bf in \N=4 SYM and the AdS/CFT correspondence} 
\vskip .6truein 
\centerline {\large J. Erdmenger\footnote{jke@mitlns.mit.edu} and
M. P\'erez-Victoria\footnote{manolo@pierre.mit.edu}}   
\end{center} 
\vspace{.3cm} 
\begin{center}
{Center for Theoretical Physics, \\
 Laboratory for Nuclear Science \\
 and Department of Physics \\
 Massachusetts Institute of Technology \\
 Cambridge, MA 02139}  
\end{center}
\vspace{1.5cm} 
 
\centerline{\bf Abstract} 
\medskip 

We show that next-to-extremal correlators of chiral primary operators
in \N=4 SYM theory do not receive quantum corrections to first order
in perturbation theory. 
Furthermore we consider next-to-extremal correlators within AdS supergravity.
Here the exchange diagrams contributing to these correlators yield
results of the 
same free-field form as obtained within field theory. This suggests that
quantum corrections vanish at strong coupling as well.

\newpage
\pagestyle{plain}
\setcounter{footnote}{0}

\footskip50pt
\parindent=12pt
\parskip=.4cm

\section{Introduction}

Among its variety of applications, the Maldacena
conjecture~\cite{Maldacena,Gubser,Witten} has been at the origin of
new results in four-dimensional quantum field theory. In particular,
it has lead to the discovery of new non-renormalization properties of
\N=4 Super Yang-Mills theory (SYM). Many of these properties refer to
correlation functions of chiral primary operators, which are scalar
composite operators in the symmetric traceless representation of the 
$SU(4)$ R-symmetry group with Dynkin labels $[0,l,0]$. 
In~\cite{FreedmanAdS}, the AdS/CFT correspondence was used to argue  
that for these operators the so-called ``extremal'' correlation
functions, in which the
conformal dimension of one of the operator equals the sum of
dimensions 
of all remaining operators, do not receive any quantum corrections. 
Thus, at least at large $N$, these correlation functions are expected
to be given exactly by their free field form, which is a product of
two-point functions.
This conjecture has been partially confirmed for any
finite $N$ by an explicit field theory calculation to first order in
perturbation theory and to leading order 
in the semiclassical expansion of any instanton
sector~\cite{Bianchi}.  The non-perturbative non-renormalization of
extremal correlators has been studied by Eden \etal\
in~\cite{West}. Using superconformal Ward identities together 
with the Grassmann and harmonic analyticity of the relevant
superfields, these authors prove that
extremal correlators can indeed be expressed as products of two-point
functions. Applying the reduction formula of~\cite{Intriligator},
which relates the derivative of an $n$-point function with respect to
the coupling to an integrated $(n+1)$-point function with an operator
insertion, 
they also show that extremal four-point correlators are independent of
the coupling constant, under the (plausible) assumption that there are
no undesired contributions from contact terms \cite{HSSW,PS}. 

Subject to the same assumption, Eden \etal\ also prove in~\cite{West} the
non-renormalization of ``next-to-extremal'' four-point correlation
functions, for which 
$k_4=\sum_{i=1}^{3} k_i - 2$, with $k_i$ the conformal dimension of
each operator. 
In this letter we show explicitly that the radiative
corrections to next-to-extremal four-point correlators of chiral primary
operators vanish to order $g^2$. This provides an independent check of
the results in~\cite{West}. We then extend our arguments to show that,
to first order in perturbation theory, next-to-extremal correlators
with an arbitrary number of points are not renormalized either. 
Furthermore, for the four-point functions, we examine also the
corresponding scattering amplitudes within AdS supergravity and show
that the exchange diagrams reduce to a free-field form. The fact 
that the quartic couplings have not been computed so far prevents us
from giving a complete proof in AdS\footnote{After
this work had been completed, a paper appeared in which the relevant
quartic couplings are evaluated~\cite{quartic}. We briefly comment on
the new results in the last section, and leave for future work the
complete computation of the next-to-extremal four-point functions in AdS
supergravity.}. 
Although the results in~\cite{West} hold for any \N=2 SCFT and for
any choice of the gauge group, here we shall stick to \N=4 SYM with
$SU(N)$ gauge symmetry.

In the case of extremal $n$-point functions (and also of general
three-point functions), there is only one $SU(4)$ invariant
contraction of 
the ``flavour'' indices, \ie, there is only one $SU(4)$ singlet in the
product of the representations of the operators involved. This fact
was used in~\cite{Bianchi} and in~\cite{FreedmanCFT} to assign to the
chiral operators a
convenient irreducible representation of the
$SU(3) \times U(1)$ subgroup of $SU(4)$ that is manifest in the \N=1
language, such that
the flavour and colour structures essentially decouple.  
However, in the case of $n$-point next-to-extremal correlators there is
more than one invariant contraction of the flavour indices in
general. 
Therefore, picking just one particular $SU(3)$
representation for each operator is not general enough. Our
strategy will be to use the full $SU(4)$ structure to constrain
the possible Wick contractions. Then we show that each of these graphs
vanishes for arbitrary $SU(3)$ representations. 

The paper is organized as follows. In Section~2 we show that
the simplest four-point next-to-extremal correlator, $\la \Tr X^4 \Tr X^2
\Tr X^2 \Tr X^2 \ra$,  is not renormalized to order $g^2$.  In
Section~3, the proof is extended to any four-point
next-to-extremal correlator, and in Section~4, to any next-to-extremal
correlator with $n$ points. In Section~5 we discuss
next-to-extremal four-point functions in AdS supergravity. Finally, 
Section~6 is devoted to conclusions, including a short discussion of
instantonic corrections.

In terms of component fields, the Lagrangian of the \N=4 theory in
Euclidean space reads:
\bea
\lefteqn{{\cal L}= \frac{1}{4} F_{\mu\nu}^2 + \frac{1}{2}
\bar{\lambda} \not \! \! D \lambda + \overline{D_\mu z^i} D_\mu z^i +
\frac{1}{2} \bar{\psi}^i \not \! \! D \psi^i} && \nn
&& \mbox{} + i \sqrt{2}g f_{abc}(\bar{\lambda}_a\bar{z}^i_bL\psi^i_a -
\bar{\psi}^i_a R z^i_b \lambda_c) -  \frac{1}{\sqrt{2}}g f_{abc}
\epsilon^{ijk} ( \bar{\psi}^i_a L z^j_b \psi^k_c + \bar{\psi}^i_a R
\bar{z}^j_b\psi^k_c ) \nn
&& \mbox{} - \frac{1}{2} g^2 (f_{abc}\bar{z}^i_bz^i_c)^2 + \frac{1}{2}
g^2 f_{abc}f_{ade} \epsilon^{ijk}\epsilon^{ilm} z^i_b z^k_c
\bar{z}^l_d\bar{z}^m_e .
\eea
$L$ and $R$ are chirality projectors.
The three complex fields $z^i~(i=1,2,3)$ are combinations of the six
fundamental real scalars of \N=4 SYM, $X^I~(I=1,\ldots,6)$:
\begin{equation}
z^i= \frac{1}{\sqrt{2}} (X^i + i X^{i+3}),~~
\bar{z}^i= \frac{1}{\sqrt{2}} (X^i - i X^{i+3}).
\end{equation}
The fields $X^I$ belong to the {\bf 6} of the $SU(4)$ R-symmetry group,
whereas $z^i$ and $\bar{z}^i$ transform in the {\bf 3} and 
${\bf \bar{3}}$, respectively, of its $SU(3)$ subgroup. 
All fields are in
the adjoint representation of the $SU(N)$ gauge group.
\begin{figure}[t]
\hspace{2.5cm}
\epsfxsize=10cm
\epsfbox{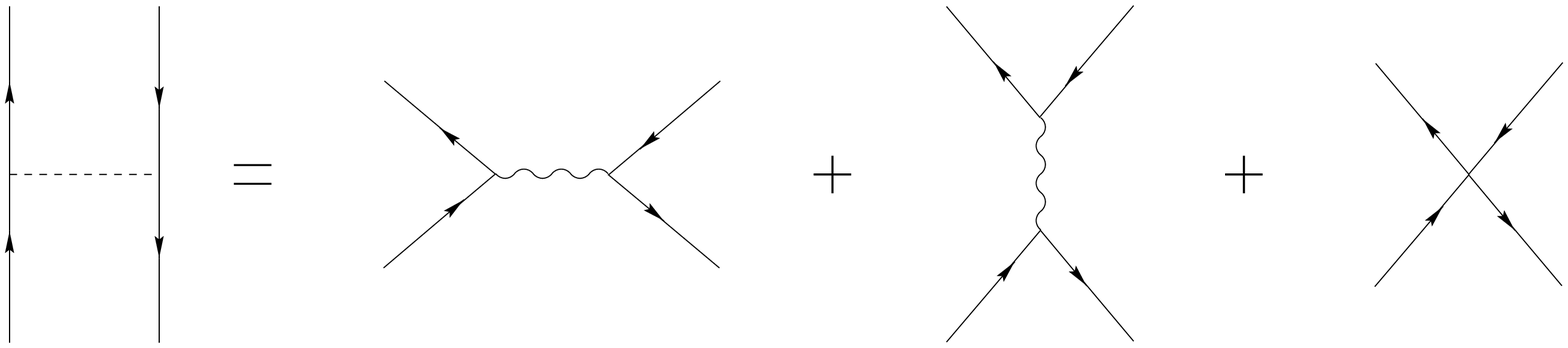}
\caption{A dashed line connecting two solid lines indicates the
effective four-scalar interaction  resulting from gluon exchange and
from the quartic vertex. Scalar propagators are denoted by solid lines
and gluon propagators by wavy lines. \label{fig1}} 
\end{figure}
It will be useful to
introduce dashed lines to represent the effective vertex obtained from
the two channels for a gluon exchange plus the four scalar contact
interaction, as indicated in Fig.~\ref{fig1}. The oriented solid lines
indicate propagators connecting a $z$ and a $\bar{z}$ field,
with the arrow going from $z$ to $\bar{z}$. On the other hand, we
shall use solid lines without arrows to represent propagators of two $X$
fields. The unoriented lines are equivalent to the sum of two
oriented lines with opposite orientations, since
$X^I(x)X^I(y)=z^i(x)\bar{z}^i(y) + \bar{z}^i(x)z^i(y)$.


\section{The correlator $\la \Tr X^4 \Tr X^2 \Tr X^2 \Tr X^2 \ra$}


\label{simplest}
We consider first, for simplicity, the order $g^2$ corrections
to the next-to-extremal correlation function $\la \Tr X^4(w) \Tr
X^2(x) \Tr X^2(y)  \Tr X^2(z) \ra$. The single-trace chiral primary
operators  $\Tr X^k\equiv \Tr X^{\{i_1}X^{i_2} \cdots X^{i_k\}}$ are
traceless symmetric tensor products of the fields $X^i$. The
operators $\Tr X^4$ and $\Tr X^2$ 
are in the {\bf 105} (with Dynkin labels $[0,4,0]$) and the
$\mbox{\bf 20}^\prime$ ($[0,2,0]$) of $SU(4)$, respectively. 
These operators can be written as
\bea
\Tr X^k & = & X_{a_1}^{\{i_1} \cdots X_{a_k}^{i_k\}} \Tr(T^{a_1}
              \cdots T^{a_k}) \nn
        & = & X_{a_1}^{<i_1} \cdots X_{a_k}^{i_k>} \Str(T^{a_1}
              \cdots T^{a_k}), 
\eea
where $X^{<i_1} \cdots X^{i_k>}$ denotes a (nonsymmetric) traceless
tensor product, $T^a$ are the hermitian generators of $SU(N)$ with
normalization $\Tr (T^a T^b) = \delta^{ab}/2$ and the symmetric trace
is defined by
\begin{equation}
\Str(T^{a_1} \cdots T^{a_k}) = \sum_{\mbox{\scriptsize perms
}\sigma} \frac{1}{k!}
\Tr(T^{a_{\sigma(1)}} \cdots T^{a_{\sigma(k)}}).
\end{equation}
Here and in the following we use the notation of
Ref.~\cite{FreedmanCFT}. 
The case of multi-trace chiral primary operators will be discussed in
the conclusions. 

\begin{figure}[ht]
\begin{center}
\hspace{1cm}
\epsfxsize=12cm
\epsfbox{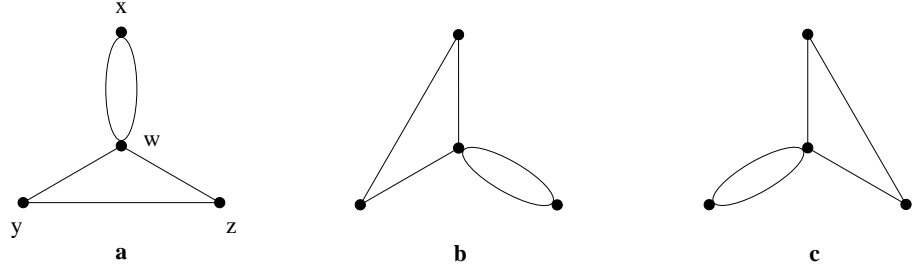}
\end{center}
\caption{Feynman diagrams contributing to the correlator $\la \Tr X^4(w)
\Tr X^2(x) \Tr X^2(y)$  $\Tr X^2(z) \ra$ at the free-field
level. \label{fig2}}  
\end{figure}
Disconnected diagrams do not contribute
because they factorize into two connected subdiagrams, one of which
($\la \Tr X^4(w) \Tr X^2(x)\ra$, for instance) cannot be a $SU(4)$
singlet.
The connected Feynman diagrams contributing at the free-field level
are depicted in Fig.~\ref{fig2}. Note that diagrams with a scalar loop
attached to a single point (a ``tadpole'') vanish due to the
tracelessness of the operator $\Tr X^4$. Diagram $a$ gives the
following contribution to the free field correlation function: 
\bea
\lefteqn{\la \Tr X^4(w) \Tr X^2(x) \Tr X^2(y)  \Tr X^2(z) \ra_a}
&& \nn
&& = {\cal C}^{I_1\ldots I_4;J_1J_2;K_1K_2;L_1L_2} 
G(w,x)^2 G(w,y) G(w,z) G(y,z) Q_{4,2,2,2}(N),
\eea
where $I_i,~J_i,~K_i$ and $L_i$ are the flavour indices of each
operator, ${\cal C}$ is a tensor in flavour space, $G(x,y)=1/(4\pi^2
(x-y)^2)$ is the scalar propagator and
\begin{equation}
Q_{4,2,2,2} = \Str(T^{a_1}\cdots T^{a_4})
\Str(T^{a_1}T^{a_2}) \Str(T^{a_3}T^b) \Str(T^{a_4} T^b)
\end{equation}
contains the colour structure. The space-time structure
factorizes into a two-point function, $G(w,x)^2$, times a three-point
function,  $G(w,y) G(w,z) G(y,z)$. 
The contributions of diagrams $b$ and $c$ are obtained by
the permutations $(x,J_i)\leftrightarrow (y,K_i)$ and
$(x,J_i)\leftrightarrow (z,L_i)$, respectively. In the future we shall
not consider the permuted diagrams explicitly, since they can be
treated in exactly the same way.  

\begin{figure}[ht]
\begin{center}
\epsfxsize=14cm
\epsfbox{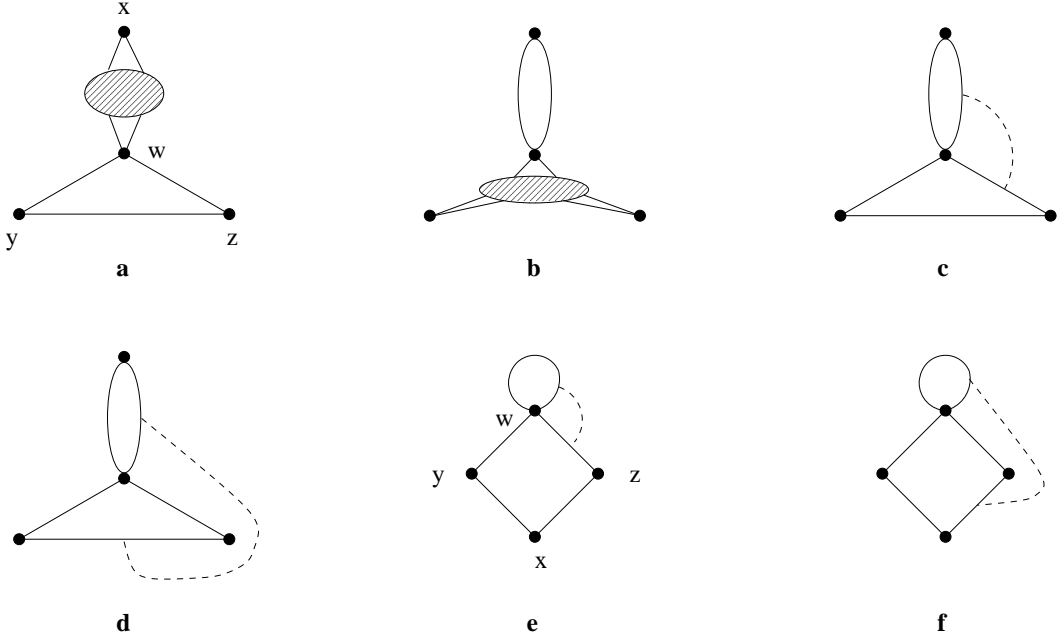}
\end{center}
\caption{Feynman diagrams contributing to the correlator $\la \Tr X^4(w)
\Tr X^2(x) \Tr X^2(y)$ $\Tr X^2(z) \ra$ at order $g^2$. We do not show
explicitly 
diagrams that are obtained from these by permutations of the
operators. In diagrams $c$ and $d$, the dashed line can be connected
to any of the two solid lines between $x$ and $w$. \label{fig3}}
\end{figure}
The diagrams contributing at order $g^2$
(up to permutations of the operators) are depicted in Fig.~\ref{fig3}.
We shall show that each of them
vanishes separately. For diagrams (actually, groups of diagrams) $a$
and $b$, this is due to known non-renormalization theorems for 2- and
3-point functions, respectively. Let us consider diagram
$b$, which gives the contribution
\bea
& {\la [X_{b_1}^{\{J_1} X_{b_2}^{J_2\}}](x) [X_{a_1}^{\{I_1}
X_{a_2}^{I_2}](w) \ra}^{(0)} 
{\la [X_{a_3}^{I_3} X_{a_4}^{I_4\}}](w) [X_{c_1}^{\{K_1}
X_{c_2}^{K_2\}}](y)  [X_{d_1}^{\{L_1} X_{d_2}^{L_2\}}](z) \ra}^{(1)}
& \nn
& \cdot~\Tr(T^{a_1} T^{a_2}T^{a_3}T^{a_4}) \Tr(T^{b_1}T^{b_2})
\Tr(T^{c_1}T^{c_2}) \Tr(T^{d_1}T^{d_2}), &  \label{two_three}
\eea
where the superindices $(0)$ and $(1)$ indicate free-field and
first ($g^2$) order correlators, respectively. Since the scalar
propagators are diagonal in flavour space,
the two flavour indices $J_1$ and $J_2$ of $\Tr X^2(x)$ in
(\ref{two_three}) are contracted with two of the
indices of $\Tr X^4(w)$, $I_1,\ldots,I_4$, within the free two-point
function. This leaves a symmetric traceless tensor of
rank 2. Therefore, as far as the flavour structure is concerned, the
operator entering the three-point function at $w$ is in the
$\mbox{\bf 20}^\prime$ of $SU(4)$. Likewise, the contraction of the
colour indices $b_1$ and $b_2$ with $a_1$ and $a_2$ within the
two-point function, and the
normalization $\Tr (T^{b_1}T^{b_2})=\delta^{b_1b_2}/2$, imply that
the colour indices of the two generators $T^{a_1}$, $T^{a_2}$ are
contracted to give the Casimir of their representation, leaving only
$T^{a_3}$ and $T^{a_4}$ in that trace. Thus, the (order $g^2$)
three-point function factor is
\begin{equation}
{\la \Tr X^2(w) \Tr X^2(y) \Tr X^2(z)\ra}^{(1)},
\label{3part}
\end{equation}
with three chiral primaries in the $\mbox{\bf 20}^\prime$. 
This expression must vanish, since it is known
that the correlator $\la \Tr (X^2)(w) \Tr X^2(y) \Tr X^2(z)\ra$ does not
receive quantum corrections. Hence, diagram $b$ vanishes.
A similar argument holds for diagram $a$, which vanishes as well.

Any of the diagrams of Fig.~\ref{fig3} can be decomposed into a sum of
``oriented'' diagrams with
arrows on the scalar lines. Different ``orientations''
correspond either to different $SU(3)$ representations in the
$SU(4)\rightarrow SU(3) \times U(1)$ decomposition of the 
operators or to different Wick contractions of a given set of
$SU(3)$ representations\footnote{The chiral
primary operators are linear combinations of $SU(3)$ irreducible
representations. For instance, if $I=i,~J=j \in \{1,2,3\}$, 
$\Tr X^{\{I}X^{J\}} = \frac{1}{2} \Tr(z^{(i}z^{j)} +
\bar{z}^{(i}\bar{z}^{j)} + z^{\{i}\bar{z}^{j\}} + \bar{z}^{\{i}
z^{j\}})$.}. The explicit colour and flavour
structure of each diagram depends on the assignment of
arrows to the scalar lines. However, all the terms in the possible
oriented diagrams
contain the structure $f_{abp} f_{cdp}$, where $a$, $b$, $c$ and $d$
are the colour indices of the four scalar lines involved in the gluon
exchange or the contact interaction. In the case of gluon exchange and
of the contact interaction coming from the $D$-term in the Lagrangian,
the colour indices $a$ and $b$ (and $c$ and $d$) are attached to one
incoming and one outgoing scalar line, while for the $F$-term
contact interaction, they are attached to two incoming or two outgoing
lines. The essential fact for the present argument is that, for fixed
colour indices in the relevant scalar propagators, there are only
three possible colour combinations: $f_{abp} f_{cdp}$, $f_{acp}
f_{bdp}$ and $f_{adp} f_{bcp}$. This structure is enough to prove that
any oriented diagram corresponding to the diagrams $c$, $d$, $e$ and
$f$ vanishes.

Let us consider diagram $c$. For any given orientation, its colour
structure is given by:
\bea
\lefteqn{\Str(T^{a_1}\cdots T^{a_4}) \Str(T^{a_1}T^{b}) \Str(T^c T^d)
\Str(T^d T^{a_4})} && \nn
&& \left( A f_{a_2bp} f_{a_3cp} + B f_{a_2cp} f_{a_3bp} +
C f_{a_2a_3p} f_{bcp} \right) \nn
&=& \frac{1}{6} \Str(T^{a_1}\cdots T^{a_4}) \cdot \left(A f_{a_2a_1p}
f_{a_3a_4p} + B f_{a_2a_4p} f_{a_3a_1p} + C f_{a_2a_3p} f_{a_1a_3p}
\right) \nn
&=& 0,  \label{diagc}
\eea
where $A$, $B$ and $C$ are functions of the points $w$, $x$, $y$ and
$z$ and of the flavour indices, and we have used
$\Tr (T^aT^b)=\delta^{ab}/2$. Expression (\ref{diagc}) vanishes since,
for each term, two colour indices of the totally symmetric
$\Str(T^{a_1}\cdots T^{a_4})$ are contracted with two indices of an
antisymmetric structure constant.
The remaining diagrams, $d$, $e$ and $f$, vanish for the same reason:
they involve a contraction of symmetric and antisymmetric tensors as
well. 

Therefore, the
correlator $\la \Tr X^4(w) \Tr X^2(x) \Tr X^2(y)  \Tr X^2(z) \ra$ does
not receive quantum corrections at order $g^2$. For this
correlation function, the proof is simplified by the fact that the
trace of $T^aT^b$ gives a Kronecker delta. In the next
section we shall see that the argument can be generalized to operators
of higher dimension.


\section{Next-to-extremal four-point correlators}

\label{fourpoints}

In this section we consider
the correlation function $\la \Tr X^{k_1}(w) \Tr
X^{k_2}(x) \Tr X^{k_3}(y) \Tr X^{k_4}(z) \ra$, with
$k_1=k_2+k_3+k_4-2$ and $k_i \geq 2$. Each operator is in the $SU(4)$
irreducible representation $[0,k_i,0]$. In order to show that its
radiative corrections vanish at first order, we shall group---as we
did in last section---all the diagrams whose space-time part
factorizes into a free two-point subdiagram times a 
three-point subdiagram at order $g^2$, or the 
other way round. Then, we shall invoke the non-renormalization theorems
proved in~\cite{FreedmanCFT} (see also~\cite{Park1}) for two- and
three-point functions to
conclude that the sum of all these diagrams cancel.
As we shall discuss, the proofs given in~\cite{FreedmanCFT}
apply also to the mentioned subdiagrams, which involve chiral primaries
with insertions of extra $SU(N)$ generators in the trace of one
of the operators. For the other diagrams, we shall see that they
vanish due to their colour structure.

Let us consider first the free-field contributions. It is impossible
to construct a 
$SU(4)$ invariant disconnected diagram, since (for $i,j,l,m$ all
different) $k_i=k_j$ and $k_l=k_m$ cannot hold at the same time. 
The next-to-extremality condition, together with tracelessness in flavour
space, puts also severe restrictions on the possible connected Feynman 
graphs. 
\begin{figure}[ht]
\begin{center}
\epsfxsize=12cm
\hspace{1cm}
\epsfbox{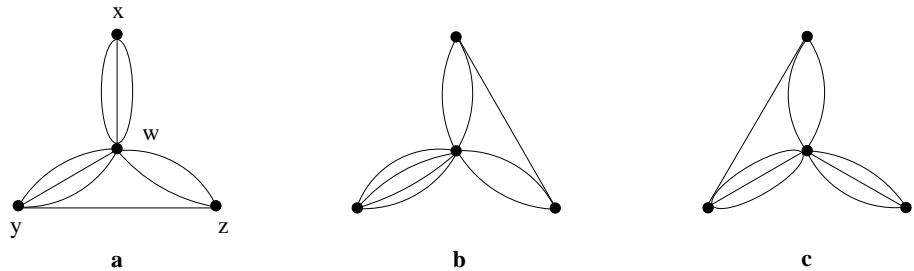}
\end{center}
\caption{Feynman diagrams contributing to the correlator $\la \Tr
X^{k_1}(w) \Tr X^{k_2}(x) \Tr X^{k_3}(y)$ $\Tr X^{k_4}(z) \ra$  at the
free-field level. Diagrams are drawn for $k_1=8$, $k_2=3$, $k_3=4$ and
$k_4=3$. For arbitrary values of $k_i$, solid lines have to be added
to or removed from the different ``rainbows''. \label{fig4}} 
\end{figure}
\newpage
\begin{figure}[thb]
\begin{center}
\epsfxsize=14cm
\epsfbox{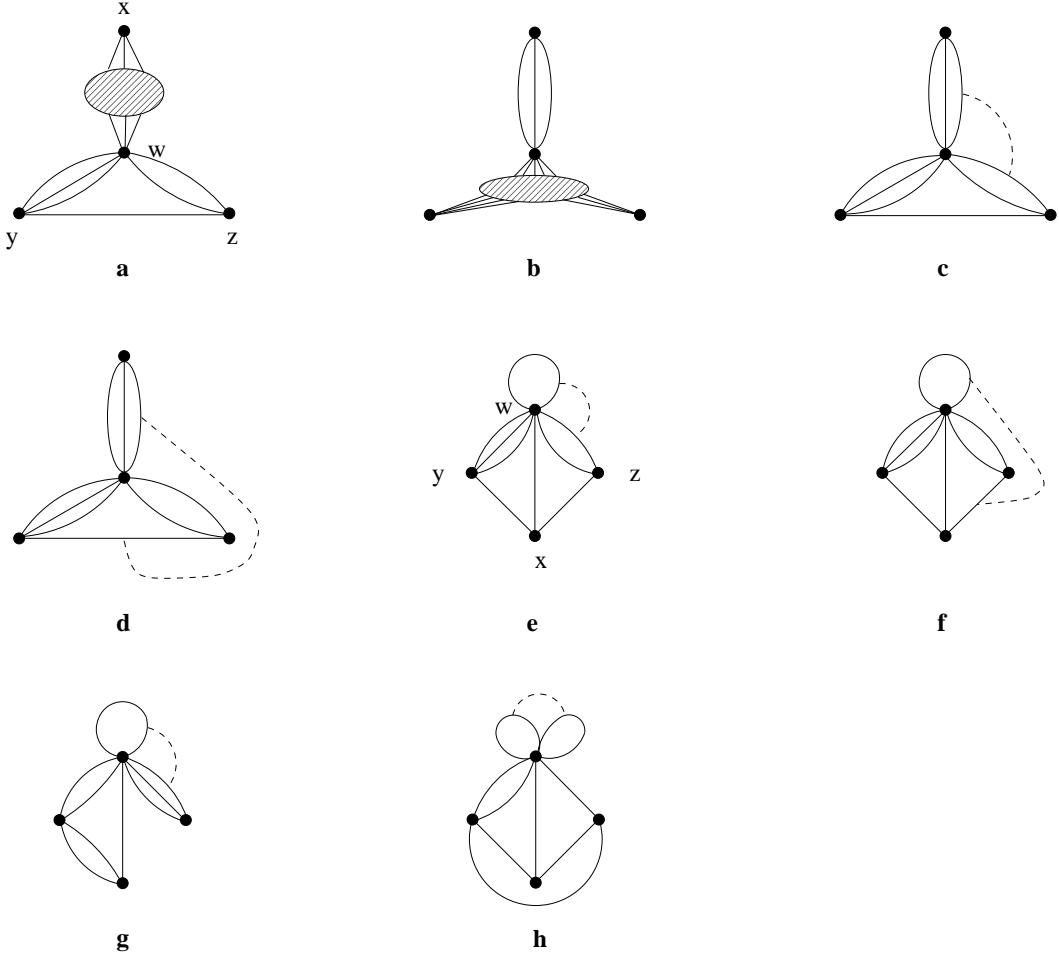}
\end{center}
\caption{Feynman diagrams contributing to the correlator $\la \Tr
X^{k_1}(w) \Tr X^{k_2}(x) \Tr X^{k_3}(y)$ $\Tr X^{k_4}(z) \ra$ at order
$g^2$. As in Fig. 4, diagrams are drawn for $k_1=8$, $k_2=3$, $k_3=4$
and $k_4=3$. The dashed lines can be connected
to any of the solid lines in each ``rainbow''.
Here and in the following, permutations are not shown
explicitly. Diagram $h$ is just one of the various diagrams that have
two tadpoles connected by a dashed
line.   \label{fig5}}
\end{figure}
The only diagrams contributing at the free-field level are
depicted in Fig.~\ref{fig4}. They all have the same structure, so we
shall focus on the first one, which yields 
\begin{equation}
{\cal C}^{I_1\ldots I_{k_1};J_1\ldots J_{k_2};K_1\ldots
K_{k_3};L_1\ldots L_{k_4}}  G(w,x)^{k_2} G(w,y)^{k_3-1} G(w,z)^{k_4-1}
G(y,z) Q_{k_1,k_2,k_3,k_4}(N), \label{freefieldresult} 
\end{equation}
where
\bea
\lefteqn{Q_{k_1,k_2,k_3,k_4} = \Str(T^{a_1}\cdots
T^{a_{k_1}}) \Str(T^{a_1}\cdots T^{a_{k_2}})} && \nn
&& \mbox{} \cdot \Str(T^{a_{k_2+1}} \cdots T^{a_{k_2+k_3-1}} T^b)
\Str(T^{a_{k_2+k_3}} \cdots T^{a_{k_1}} T^b).
\eea
We see that in the free-field approximation the space-time structure
factorizes into a
two-point function and a three-point function. The two-point
function consists of a ``rainbow'' of scalar lines connecting $w$ and
$x$, while the three-point function has two ``rainbows'' connecting
$w$ to $y$ and $z$, and one additional scalar line between $y$ and
$z$.

At order $g^2$, the possible kinds of diagrams are depicted 
in Fig.~\ref{fig5}. In addition to the diagrams we show, there are
others obtained by permutations of the operators at $x$, $y$, and $z$,
and by attaching the dashed line to different scalar lines in the
``rainbows''. It will be sufficient to study the set of
diagrams in Fig.~6, for the remaining diagrams can be shown to vanish
in exactly the same manner.
As in the case of the correlator $\la \Tr X^4 \Tr X^2 \Tr X^2 \Tr X^2
\ra$, diagrams $a$ and $b$ only contain radiative corrections to a
two-point and a three point function, respectively. In this
case their vanishing follows from a simple generalization of the proof
in~\cite{FreedmanCFT} of the non-renormalization of general correlators
of two or three chiral primaries. 

Let us start with diagram $a$, which gives the contribution
\bea
&&\lefteqn{{\la [X_{b_1}^{\{J_1} \cdots X_{b_{k_2}}^{J_{k_2}\}}](x) 
[X_{a_1}^{\{I_1} \cdots X_{a_{k_2}}^{I_{k_2}}](w) \ra}^{(1)}} \nn
& \mbox{} \cdot & {\la [X_{a_{k_2+1}}^{I_{k_2+1}} \cdots
X_{a_{k_1}}^{I_{k_1}\}}](w) [X_{c_1}^{\{K_1} \cdots 
X_{c_{k_3}}^{K_{k_3}\}}](y)  [X_{d_1}^{\{L_1} \cdots
X_{d_{k_4}}^{L_{k_4}\}}](z) \ra}^{(0)} \nn
& \mbox{} \cdot & \Tr(T^{a_1} \cdots T^{a_{k_1}}) \Tr(T^{b_1} \cdots
T^{b_{k_2}}) \Tr(T^{c_1} \cdots T^{c_{k_3}}) \Tr(T^{d_1}\cdots
T^{d_{k_4}}). 
\eea
The operator entering the two point function at $w$ is a
symmetric tracessless tensor in flavour space, so it belongs to the
$[0,k_2,0]$ representation of $SU(4)$. Therefore we are lead to
consider the two-point function of two chiral primaries of conformal
dimension $k_2$ in the representation $[0,k_2,0]$, but with $k_1-k_2$
additional $SU(N)$ generators inserted in one of the traces:
\begin{equation}
\la \Tr X^{k_2}(x) \Tr (X^{k_2}(w) T^{a_{k_2+1}} \cdots
T^{a_{k_1}}) \ra . \label{2pf}
\end{equation}
The next step is to realize
that the proof in~\cite{FreedmanCFT} of the non-renormalization of the
correlator $\la \Tr X^k \Tr X^k \ra$ applies also to the
two-point correlator (\ref{2pf}).
Indeed, it is clear that self-energy corrections are not affected by
the extra generators in the second trace. On the other hand, the pairs
of structure constants appearing 
in gluon exchanges and quartic vertices can be converted into
commutators within the {\em first} trace. The sum of all these
diagrams can then be reduced to $k_2$ identical terms, as
in Eqs.~(2.5) and~(2.6) of~\cite{FreedmanCFT}. The key point is that
there are not additional factors in the first trace that would give
extra terms upon the application of the identity
\begin{equation}
\sum_{i=1}^n \Tr(M_1 \cdots [N,M_i]\cdots M_n)=0. \label{traces}
\end{equation}
Note that the presence of extra generators inside both traces would
prevent us from using this argument. Therefore, after adding all
contributions one finds that the radiative corrections to the
correlator in Eq.~(\ref{2pf}) are of the form of Eq.(2.7)
in~\cite{FreedmanCFT}, and hence the non-renormalization theorem
for $\la \Tr X^2 \Tr X^2 \ra$ implies that they vanish.

In the same manner, the non-renormalization theorems for three-point
functions imply that diagram $b$ vanishes. Here, we must consider the
correlator
\begin{equation}
\la \Tr (T^{a_1} \cdots T^{a_{k_2}} X^{k_1-k_2}(w)) \Tr X^{k_3}(y) \Tr
X^{k_4}(z) \ra . \label{3pf}
\end{equation}
Self-energy corrections are again trivial, and
interactions relating scalar lines within one ``rainbow'' can be
treated as in Eq.~(4.7) of~\cite{FreedmanCFT} if one chooses
to put the commutators inside the second or third trace, which is
always possible. Furthermore, one can easily see that the
manipulations carried out in~\cite{FreedmanCFT} for diagrams with
interactions 
among two different  ``rainbows'' can also be done in our case. 
Indeed, as long as only one of the traces contains extra $T$'s, there
is no obstruction to the use of Eq.~(\ref{traces}). Therefore, the sum
of all contributions is of the form of Eq.~(4.9) of~\cite{FreedmanCFT}
and vanishes due to the non-renormalization theorems for $\la \Tr
X^2 \Tr X^2 \ra$ and $\la \Tr X^2 \Tr X^2 \Tr X^2 \ra$.

Let us study now the genuine four-point diagrams. Each of them is a
linear combination of oriented diagrams. The general color structure
of the interactions, which was discussed in last section, will be
sufficient again to show that all possible oriented diagrams vanish. 
Consider diagram $c$ of Fig.~\ref{fig5}. For any given orientation, it
gives a contribution of the following form:
\bea
\lefteqn{\Str(T^{a_1}\cdots T^{a_{k_1}})   \Str(T^{a_1} \cdots
T^{a_{k_2-1}} T^b)   \Str(T^c T^{a_{k_2+2}} \cdots
T^{a_{k_2+k_3-1}} T^d)  } && \nn  
&& \mbox{} \cdot \Str(T^d T^{a_{k_2+k_3}} \cdots T^{a_{k_1}}) 
\left( A^\prime f_{a_{k_2}bp} f_{a_{k_2+1}cp} + B^\prime f_{a_{k_2}cp}
f_{a_{k_2+1}bp} + C^\prime f_{a_{k_2}a_{k_2+1}p} f_{bcp} \right),
\eea
where $A^\prime$, $B^\prime$ and $C^\prime$ carry the space-time and
flavour dependence. The term involving $C^\prime$ vanishes because
$f_{a_{k_2}a_{k_2+1}p}$ 
is contracted to a trace that is symmetric in $a_{k_2},a_{k_2+1}$. We
convert one of the structure constants in the terms involving
$A^\prime$ ($B^\prime$) into a commutator inside the second trace, and
then use Eq.~(\ref{traces}) to obtain
\bea
\lefteqn{i\Str(T^{a_1}\cdots T^{a_{k_1}}) 
\Str(T^c T^{a_{k_2+2}} \cdots
T^{a_{k_2+k_3-1}} T^d) \Str(T^d T^{a_{k_2+k_3}} \cdots T^{a_{k_1}}) }
&& \nn
&& \mbox{} \cdot  \left\{ A^\prime f_{a_{k_2+1}cp}
\Str(T^{a_1} \cdots T^{a_{k_2-1}} [T^{a_{k_2}},T^p])
+ B^\prime f_{a_{k_2}cp} \Str(T^{a_1} \cdots T^{a_{k_2-1}}
[T^{a_{k_2+1}},T^p]) \right\}  \nn
&=& - i\Str(T^{a_1}\cdots T^{a_{k_1}})   \Str(T^c T^{a_{k_2+2}} \cdots
T^{a_{k_2+k_3-1}} T^d)   \Str(T^d T^{a_{k_2+k_3}} \cdots
T^{a_{k_1}}) \nn 
&& \mbox{} \cdot  \sum_{i=1}^{k_2-1} \left\{ A f_{a_{k_2+1}cp}
\Str(T^{a_1} \cdots [T^{a_{k_2}},T^{a_i}] \cdots T^{a_{k_2-1}} T^p)
\right. \nn
&& \left. \mbox{} + B f_{a_{k_2}cp} \Str(T^{a_1} \cdots
[T^{a_{k_2+1}},T^{a_i}] \cdots  T^{a_{k_2-1}} T^p) \right\}.
\eea
The traces must not be symmetrized in the generators arising
from the structure constants.
Each term in the sum contains a commutator that is antisymmetric in
$a_{k_2},a_i$ or in $a_{k_2+1},a_i$. Since the sum is contracted with
a trace symmetric in those indices, the entire diagram vanishes.

Diagram $d$ gives the contribution
\bea
\lefteqn{\Str(T^{a_1}\cdots T^{a_{k_1}})   \Str(T^{a_1} \cdots
T^{a_{k_2-1}} T^b)   \Str(T^{a_{k_2+1}} \cdots
T^{a_{k_2+k_3-1}} T^c)  } && \nn  
&& \mbox{} \cdot \Str(T^d T^{a_{k_2+k_3}} \cdots T^{a_{k_1}}) 
\left( A^{\prime\prime} f_{a_{k_2}bp} f_{cdp} + 
B^{\prime\prime} f_{a_{k_2}cp}
f_{dbp} + C^{\prime\prime} f_{a_{k_2}dp} f_{bcp} \right).
\eea
For the terms involving $A^{\prime\prime}$, $B^{\prime\prime}$
and $C^{\prime\prime}$, the structure constant with the index $a_{k_2}$
can be converted into a commutator inside the second, third and fourth
trace, respectively. The identity~(\ref{traces}) then leads to a sum
of terms with commutators $[T^{a_{k_2}},T^{a_i}]$ that cancel when
contracted with the first symmetric trace. Let us show this explicitly
for the term involving $B^{\prime\prime}$:
\bea
\lefteqn{i B^{\prime\prime} f_{dbp} \Str(T^{a_1}\cdots T^{a_{k_1}})  
\Str(T^{a_1} \cdots 
T^{a_{k_2-1}} T^b) } && \nn 
&& \mbox{} \cdot \Str(T^{a_{k_2+1}} \cdots
T^{a_{k_2+k_3-1}} [T^{a_{k_2}},T^p])  
\Str(T^d T^{a_{k_2+k_3}} \cdots T^{a_{k_1}}) \nn
&=& - i B f_{dbp} \Str(T^{a_1}\cdots T^{a_{k_1}}) \Str(T^{a_1} \cdots
T^{a_{k_2-1}} T^b) \nn 
&& \mbox{} \cdot \sum_{i=k_2+1}^{k_2+k_3-1} \Str(T^{a_{k_2+1}} \cdots
[T^{a_{k_2}},T^{a_i}] \cdots T^{a_{k_2+k_3-1}} T^p)
\Str(T^d ) T^{a_{k_2+k_3}} \cdots T^{a_{k_1}}) \nn
&=& 0
\eea
It is interesting to note that this argument would not hold if there
were another scalar line connecting the operators at $y$ and $z$. Such
a diagram is forbidden by next-to-extremality, but it can contribute to
next-to-next-to-extremal correlators.

The remaining diagrams have a dashed line connected to a tadpole. All
diagrams with tadpoles must have interactions involving all the
tadpole lines, for otherwise they would vanish due to the
tracelessness of the operators. Diagrams $e$, $g$ and
$h$~of Fig.~6 vanish because all possible terms contain the factor
\begin{equation}
f_{a_ia_jp} \Str(T^{a_1}\cdots T^{a_{k_1}}) = 0, \label{trivial}
\end{equation}
where $i=1,2$, $j=1,\ldots,k_1$
Finally, diagram $f$ has three possible terms, one proportional to the
color structure in Eq.~(\ref{trivial} ) and two to a factor of the
form ($i=1,2$):
\bea
\lefteqn{f_{a_ibp} \Str(T^{a_1} \cdots T^{a_{k_1}}) \Str(T^bT^{a_{k_2+1}}
\cdots T^{a_{k_2+k_3-1}})} && \nn
&=& i \Str(T^{a_1} \cdots T^{a_{k_1}}) \Str([T^{a_i},T^p]T^{a_{k_2+1}}
\cdots T^{a_{k_2+k_3-1}}) \nn
&=& -i \Str(T^{a_1} \cdots T^{a_{k_1}}) \sum_{j=k_2+1}^{k_2+k_3-1}
\Str([T^p T^{a_{k_2+1}} \cdots [T^{a_i},T^{a_j}] \cdots
T^{a_{k_2+k_3-1}}) \nn 
&=& 0.
\eea

\begin{figure}[ht]
\hspace{6cm}
\epsfxsize=3cm
\epsfbox{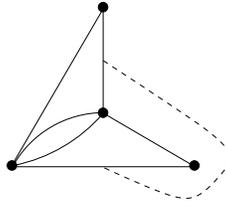}
\caption{A Feynman diagram contributing to a next-to-next-to-extremal
correlator at order $g^2$. \label{fig6}}  
\end{figure}
Therefore, we conclude that the correlator $\la \Tr X^{k_1}(w) \Tr
X^{k_2}(x) \Tr X^{k_3}(y) \Tr X^{k_4}(z) \ra$ does not receive quantum 
corrections to order $g^2$. On the other hand, in the case of
four-point correlators that are neither extremal nor next-to-extremal,
non-vanishing Feynman diagrams (as the one illustrated in
Fig.~\ref{fig6}) contribute at order $g^2$, and there is no apparent
reason for them to cancel. Therefore we do not expect any
non-renormalization theorem for $k_1<k_2+k_3+k_4-2$, in agreement with
the argument of~\cite{West}. An example for this is provided by the
explicit calculation of $g^2$ corrections to $\la \Tr X^2(w) \Tr
X^2(x) \Tr X^2(y) \Tr X^2(z) \ra$ in~\cite{Park2}.


\section{General next-to-extremal correlators}


The arguments given in last section can be directly extended to a
general $n$-point function, $\la \Tr X^{k_1}(x_1) \Tr
X^{k_2}(x_2) \cdots \Tr X^{k_n}(x_n) \ra$, with $k_1=\sum_{i=2}^{n}
k_i-2$ and $k_i \geq 2$.  As in the case of four-point next-to-extremal
correlators, disconnected diagrams are forbidden by $SU(4)$
symmetry. The reason is that the next-to-extremality condition implies that
$k_1 > \sum_{i=1}^n k_i - k_l - k_m$ for any $l,m$, so that if two or
more operators form a subdiagram disconnected from the rest
of the diagram, it is impossible to build a $SU(4)$ invariant
subdiagram with the remaining operators.

\begin{figure}[ht]
\hspace{6cm}
\epsfxsize=4cm
\epsfbox{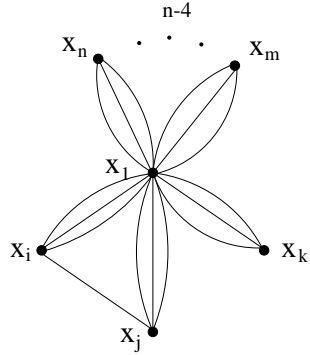}
\caption{Feynman diagrams contributing to an $n$-point
next-to-extremal correlator at the free-field level. \label{fig7}} 
\end{figure}
\begin{figure}[ht]
\hspace{2.5cm}
\epsfxsize=10cm
\epsfbox{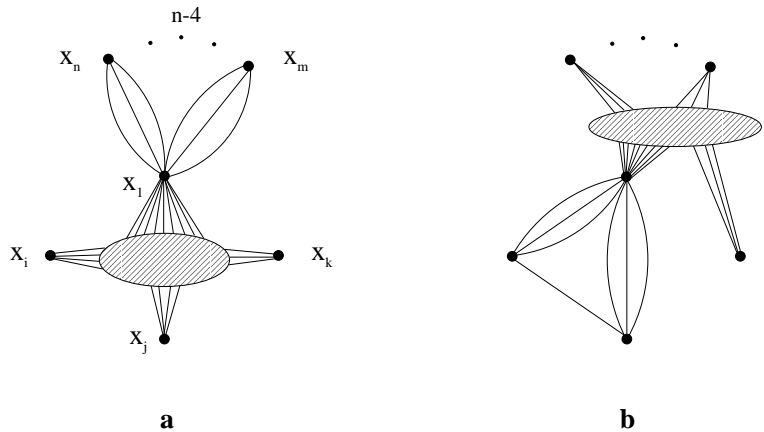}
\caption{Feynman diagrams contributing to an $n$-point
next-to-extremal correlator at order $g^2$. \label{fig8}}
\end{figure}
The free-field connected Feynman diagrams (Fig.~\ref{fig7}) are
analogous to the corresponding ones for four-point correlators: their
space-time part factorizes into a product of $(n-3)$ two-point
functions 
times one three-point function. A similar structure is
preserved at order $g^2$: as is shown in Fig.~\ref{fig8}, the
space-time part 
of all diagrams factorizes either into a free extremal function of
$(n-4)$ points times a first-order four-point next-to-extremal function
(diagram $a$) or into a free three-point next-to-extremal
function times a first-order extremal function of $(n-3)$ points
(diagram $b$).

In diagram $a$, the operator that enters the four-point function at
$w$ is a traceless symmetric tensor of rank $k_i+k_j+k_l-2$, where
$i,j,l \in \{2,\ldots,n\}$ label the other three
operators involved in the four-point function.
Hence, we are dealing with a next-to-extremal four-point correlator of the
form 
\begin{equation}
\la \Tr(T^{a_1}\cdots T^{a_{k_1-k_i-k_j-k_l+2}} X^{k_i+k_j+k_l-2}(x_1)) 
\Tr X^{k_i}(x_i) \Tr X^{k_j}(x_j) \Tr X^{k_l}(x_l) \ra
\label{fourpointsub}
\end{equation}
We have proved in Section~\ref{fourpoints} that
next-to-extremal four-point  correlators of chiral primaries are not
renormalized, and we only 
need to make sure that the argument also holds in the presence of
additional $SU(N)$ generators inside the first trace. This is indeed
the case since the extra generators are included in the
trace of the 
operator with highest dimension. Therefore, only one of the traces in
the two- and three-point subdiagrams contributing to the 
correlator~(\ref{fourpointsub}) contains extra generators, as is
necessary for the non-renormalization of these subdiagrams.
Furthermore, the diagrams that cannot be factorized into
a two-point and a three-point part vanish as well since the property
of the highest-dimension operator essential in the proof is that it is
totally symmetric in the colour indices, and this is also true in the
presence of additional generators. Therefore, next-to-extremal
four-point correlators with extra group generators attached to the
highest-dimension operator do not receive quantum corrections either.

Similarly, diagram $b$ involves an extremal correlator of $n-3$ chiral
primaries, with extra $SU(N)$ generators attached to the
highest-dimension operator:
\bea
\la \Tr(T^{a_1}\cdots T^{a_{k_i+k_j-2}} X^{k_1 - k_i - k_j + 2}(x_1))
\Tr X^{k_2}(x_2) \cdots \Tr X^{k_{i-1}}(x_{i-1}) \nn
\mbox{} \cdot \Tr X^{k_{i+1}}(x_{i+1}) \cdots \Tr X^{k_{j-1}}(x_{j-1}) 
\Tr X^{k_{j+1}}(x_{j+1}) \cdots \Tr X^{k_n}(x_n) \ra. 
\label{n_3pointsub} 
\eea
Here, $i$ and $j$ label the operators in the free-field three-point
subdiagram. The extremal correlator in~(\ref{n_3pointsub}) can be
shown to vanish using methods similar to the ones in
Section~\ref{fourpoints}. Alternatively, we may observe that the proof
in~\cite{Bianchi} of the non-renormalization of extremal $n$-point
correlators at order $g^2$ holds also in the presence of extra
generators. The arguments in~\cite{Bianchi} are based on the
space-time structure of the diagrams (which is colour independent) and
on the symmetry of the colour indices in the highest-dimension
operator, which is not affected by the extra generators.

Thus, we see that the combination of non-renormalization theorems for
extremal $n$-point functions and for next-to-extremal four-point
functions implies the vanishing of any next-to-extremal $n$-point
function.

\section{Next-to-extremal correlators within AdS super\-gravity}

Further evidence for the non-renormalization of next-to-extremal 
correlation functions may be obtained from calculations in type IIB
classical supergravity on $AdS_5\times S_5$
which, according to the AdS/CFT correspondence, is dual to \N=4 SYM at
strong coupling and large $N$. 

The supergravity states corresponding to chiral primary operators
of \N=4 SYM are certain scalar mixtures, $s_k$, of
the trace of the graviton on $S_5$ and the five form field on
$S_5$. These one-particle states have the correct transformation
properties under the superconformal group. They belong to the 
$[0,k,0]$ representation of $SU(4)$ and have conformal dimension
$\Delta=k$. In~\cite{Seiberg}
the three-point amplitudes of these fields have been
calculated and shown to be equal to their
free-field values, in agreement with the field theory
results~\cite{FreedmanCFT,Howe}. In~\cite{FreedmanAdS} 
it has been argued that 
extremal $n$-point functions satisfy non-renormalization theorems as
well, such 
that to all orders in perturbation theory they are given by a product
of two-point functions whose 
coefficient does not receive quantum corrections either. Here 
we are interested in next-to-extremal functions of four chiral primary
scalars, $s_{k_1}(x_1),\ldots,s_{k_4}(x_4)$, with
$k_1=k_2+k_3+k_4-2$.

We use the methods for calculating correlation functions in AdS space
developed in \cite{Freedman3pt}, \cite{Freedmanscalar4pt}, \cite{wrt} 
and consider the Euclidean continuation of $AdS_5$ whose metric is
given by
\begin{equation}
ds^2 = \frac{1}{{z_0}^2} ( d{z_0}^2 + \sum\limits_{i=1}^{4} d{z_i}^2 )
\, .
\end{equation}
The scalar bulk to boundary propagator is given by \cite{Witten},
\cite{Freedman3pt} 
\begin{equation}
K_\Delta (x,z) \equiv K_\Delta (0,\vec{x},z_0, \vec{z}) = 
C_\Delta \left( \frac{z_0}{{z_0}^2 + (\vec{z} - \vec{x})^2}
\right)^\Delta ,  \; C_\Delta=
\frac{\Gamma(\Delta)} {\pi^2 \Gamma(\Delta - 2)} \, .
\end{equation}
The bulk propagator $G_\Delta(z,y)$ depends on the chordal distance $u =
2z_0 y_0/ (z-y)^2$. Its explicit form is not needed here.

\begin{figure}[ht]
\hspace{2.5cm}
\epsfxsize=8cm
\epsfbox{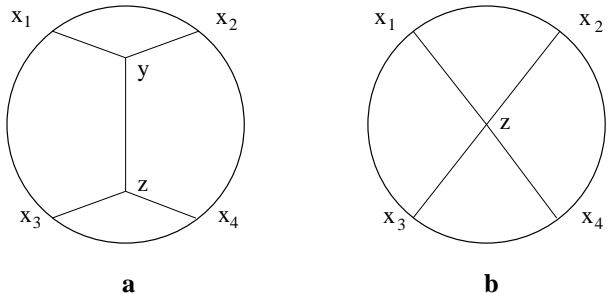}
\caption{Witten diagrams contributing to next-to-extremal four-point
functions at large $N$. To diagram $a$ one has to add the two crossed
channels for particle exchange in the bulk. \label{fig9}}
\end{figure}
We calculate the spatial dependence of the exchange diagrams 
contributing to next-to-extremal correlators by
adapting an argument given in \cite{FreedmanAdS} for the extremal
case. 
The two types of connected Witten diagrams contributing to
four-point functions are depicted in Fig.~\ref{fig9}. 
The quartic supergravity couplings required for the
evaluation of the contact diagram are not yet known, so we concentrate
on the exchange diagrams. Without loss of generality we consider
the channel where $s_{k_1}(x_1)$ and $s_{k_2}(x_2)$ join at
$y$ to the intermediate field $\phi$, which then splits at $z$ into 
the other two fields, $s_{k_3}(x_3)$ and $s_{k_4}(x_4)$, as shown in
Fig.~\ref{fig9}. The
exchanged field must be in
a common $SU(4)$ representation of the tensor products
$[0,k_1,0]\times [0,k_2,0]$ and $[0,k_3,0]\times [0,k_4,0]$. It is easy
to see using Young tableaux that the only common representations
when $k_1=k_2+k_3+k_4-2$ are $[0,k_3+k_4-2,0]$, $[0,k_3+k_4,0]$
and $[1,k_3+k_4-2,1]$. The exchanged field must be in one of these
representations, and it can be either a primary field or a $SU(2,2|4)$
descendent. 

For scalar primary exchange there are two possibilities:
$[0,k_3+k_4-2,0]$ and $[0,k_3+k_4,0]$.
The exchanged field has dimension $k_3+k_4-2$ or $k_3+k_4$,
respectively. In the first case, the cubic vertex at 
$y$ is extremal and the vertex at $z$ is next-to-extremal, while in
the second case the 
vertex at $y$ is next-to-extremal and the vertex at $z$ is
extremal. It is known from supergravity~\cite{Seiberg,Lee} that the 
cubic couplings, denoted by ${\cal G}(k_1,k_2,k)$,
vanish at extremality while
they are finite in the subextremal case. On the other hand,
divergent integrals appear in extremal correlators. We shall
regularize them by analytic continuation in the conformal
dimensions~\cite{analytic,FreedmanAdS}. It turns out that the zeros in
the couplings combine with the poles in the integrals to give a finite
result. 

The exchange of the primary $s_{k_3+k_4-2}$ leads to the contribution
\begin{eqnarray} 
\lefteqn{I_1 =
{{\cal G}(k_1,k_2,k_3+k_4-2)} {{\cal G}(k_3,k_4,k_3+k_4-2)}} && \nn
&& \mbox{} \cdot
\int \!\!\! \int \,  \frac{d^5 \! y}{{y_0}^5} \frac{d^5 \! z}{{z_0}^5}
\, K_{\Delta_1}(x_1, y)  K_{\Delta_2}(x_2, y)
G_{\Delta_3+\Delta_4-2}(y,z) K_{\Delta_3}(x_3, z) K_{\Delta_4}(x_4, z)
. \label{int1}
\end{eqnarray}
The evaluation of this integral is based on realizing that the
$y$-integrand, which leads to a pole since the vertex at $y$ is
extremal, is dominated by the contribution when $y \sim x_1$. 
For $y \sim x_1$ we may approximate the bulk propagator 
by \cite{Freedmanscalar4pt,Rastelli}
\begin{equation}
G_{\Delta} \longrightarrow \frac{1}{2 \Delta -d} \, 
{y_0}^\Delta K_\Delta(x_1,z) \, , \label{approx}
\end{equation}
such that the two integrals decouple and we have
\begin{gather}
I_1 = I_y \cdot I_z,
\end{gather}
where
\begin{align}
I_y & = {\cal G} (k_1,k_2,k_3+k_4-2) \frac{1}{{x_{12}}^{2 \Delta_2}} 
\int_{\cal R} \frac{d^5y}{{y_0}^5} \,
\frac{{y_0}^{\Delta_1 + \Delta_2+\Delta_3+\Delta_4 -2}}{ ({y_0}^2 +
(\vec{y} - \vec{x_1})^2)^{\Delta_1}} \, , \\
I_z &=  {\cal G} (k_3,k_4,k_3+k_4-2)  
\int \!\! \frac{d^5z}{{z_0}^5}\, K_{\Delta_3 + \Delta_4 -2} (x_1,z)
K_{\Delta_3} (x_3,z)K_{\Delta_4} (x_4,z) \, .
\end{align}
Here, $\cal R$ is a five-dimensional neighbourhood of $x_1$
for which the approximation~(\ref{approx}) is valid. We shall take
${\cal R}= \{ y\in AdS_5, y_0 < a, (\vec{y}-\vec{x}_1)^2 < a^2 \}$
for simplicity, but the result does not depend on the shape or size of
this region. 
$I_z$ is a next-to-extremal finite three point function which may
be evaluated  using the methods of 
\cite{Freedman3pt} to give
\begin{equation}
I_z \sim  \frac{1}{{x_{34}}^2 {x_{13}}^{2(\Delta_3-1)} 
{x_{14}}^{2(\Delta_4-1)}}
\, .
\end{equation}
For the evaluation of $I_y$ we use analytic continuation such that
$\delta \equiv k_2+k_3+k_4 -2 - k_1 \geq 0$, which implies
$ {\cal G} (k_1,k_2,k_3+k_4-2+\delta) \propto \delta$. Then, by
translating 
the integration variable $\vec{y} \rightarrow \vec{y} + \vec{x}_1$,
and by rescaling $\vec{y} = y_0 \vec{w}$, we obtain
\begin{align}
I_y & \sim  \delta \frac{1}{{x_{12}}^{2 \Delta_2}} 
\int_0^a \!\! \frac{d y_0}{{y_0}} {y_0}^\delta
\int_{B_{(a/y_0)}} 
\frac{d^4w}{(1 +  \vec{w}^2)^{\Delta_1}} \, ,
\end{align}
where $B_r$ is the four-dimensional ball of radius $r$ around
$\vec{w}=0$. We have
\begin{equation}
\int_{B_{r}} 
\frac{d^4w}{(1 +  \vec{w}^2)^{\Delta_1}} =
\frac{\pi^2}{(\Delta_1-2)(\Delta_1-1)} \left(1 +
\frac{r^4 (1-\Delta_1-r^{-2})(1+r^{-2})}{(1+r^2)^{\Delta_1}} \right)
\, , 
\end{equation}
such that
\begin{align}
I_y & \sim \delta \frac{1}{{x_{12}}^{2 \Delta_2}}  
\int_0^a \!\! \frac{d y_0}{{y_0}} {y_0}^\delta \left(1+O(y_0)\right)
\nn  
& \sim \delta \frac{1}{{x_{12}}^{2 \Delta_2}}
\left(\frac{a^\delta}{\delta}+O(\delta^0)\right) \, . 
\end{align}
We see that the first term in the $y_0$-integral gives a pole in
$\delta$ which is exactly cancelled 
by the factor $\delta$ arising from the coupling, and that the result
is independent of $a$ in the limit $\delta \rightarrow 0$. 
Therefore the final result for $I_1$ as given by~(\ref{int1}) is of
the form
\begin{gather}
I_1 \sim \frac{1}{{x_{21}}^{2 \Delta_2} {x_{34}}^2
{x_{13}}^{2(\Delta_3 -1)} {x_{14}}^{2(\Delta_4 -1)}} \, ,
\end{gather}
which agrees with the field-theoretical free-field result of
Eq.~(\ref{freefieldresult}).

The contribution from the exchange of the primary $s_{k_3+k_4}$ is
\begin{eqnarray} I_2 =
\lefteqn{{\cal G}(k_1,k_2,k_3+k_4) {{\cal G}(k_3,k_4,k_3+k_4)}}
&& \nn
&& \mbox{} \cdot 
\int \!\!\! \int \, \frac{d^5 \! y}{{y_0}^5} \frac{d^5 \! z}{{z_0}^5} 
\, K_{\Delta_1}(x_1, y)  K_{\Delta_2}(x_2, y)
G_{\Delta_3+\Delta_4}(y,z) K_{\Delta_3}(x_3, z) K_{\Delta_4}(x_4, z)
\, . \label{int2}
\end{eqnarray}
This vanishes since the integral is finite and ${{\cal
G}(k_3,k_4,k_3+k_4)}=0$. To see that the integral is finite we
evaluate the $y$-integral 
by applying the method of \cite{wrt} to the next-to-extremal
vertex at $y$ which yields
 \begin{align} I_2 = &\,
{{\cal G}(k_1,k_2,k_3+k_4)} {{\cal G}(k_3,k_4,k_3+k_4)}
 \nonumber\\
& \mbox{} \cdot
\sum\limits_{k=1}^{\Delta_2-1} \, a_k \, |x_{12}|^{-2\Delta_2 +2k}
 \int  \frac{d^5 \! z}{{z_0}^5}
\, K_{\Delta_1-\Delta_2 + k}(x_1, z)  K_{k}(x_2, z)
 K_{\Delta_3}(x_3, z) K_{\Delta_4}(x_4, z)
\, , \label{int3}
\end{align}
with finite coefficients $a_k$. All integrals in the sum are finite as
well.

The descendent exchange graphs are more involved, and we do not
study them in full detail here, but give an argument from which we
expect that descendent exchange does not contribute to
next-to-extremal correlators. The superconformal descendent field
$\phi$ can be a scalar or a tensor and must be in any of the three
allowed representations: $[0,k_3-k_4-2,0]$, $[0,k_3-k_4,0]$ or
$[1,k_3-k_4,1]$. Hence, it 
has a conformal dimension $\Delta \geq \Delta_3+\Delta_4$ and is a
descendent of 
a chiral primary in the $[0,k,0]$ representation, with $k \geq
k_3+k_4$. The coupling constants for the cubic vertices $\phi s_{k_1}
s_{k_2}$ 
and $\phi s_{k_3} s_{k_4}$~\cite{Lee} are related to
primary vertices by supersymmetry. The vertex $s_k s_{k_3} s_{k_4}$ is
forbidden by $SU(4)$
symmetry if $k>k_3+k_4$. Thus the only possibility is
$k=k_3+k_4$, which corresponds to an extremal vertex with a vanishing
coupling constant. Therefore, the
corresponding coupling for the vertex $\phi s_{k_3} s_{k_4}$ must also
be zero. On the other hand, the vertex $s_k s_{k_1} s_{k_2}$ is
next-to-extremal 
for $k=k_3+k_4$. Since the conformal dimension of the field $\phi$
satisfies the relation $\Delta \geq \Delta_3+\Delta_4$, these diagrams
involve integrals similar to $I_2$ in Eq. (\ref{int2}). These
integrals must be finite
independently of the spin of the intermediate field, as
the short distance behaviour of the bulk-to-bulk propagator is
universal. We conclude
that diagrams with exchange of descendents do not contribute to
next-to-extremal functions.

Finally, for a complete argument showing that the AdS calculation
yields the free-field 
result, it is necessary to calculate the contribution from the contact
diagram. This is a 
next-to-extremal four-point scalar function, which is known to be
finite. Since in this case the entire integration region contributes,
and not just a small region  
surrounding one of the external points, 
logarithms are expected to appear and there seems to
be no reason 
for these logarithms to cancel such as to give  a product
of two-point functions. Therefore, the quartic couplings of the
primary fields, ${\cal G}(k_2+k_3+k_4-2,k_2,k_3,k_4)$, has to  vanish
if a free-field form is to be obtained. In the 
case corresponding to the correlator considered in
Section~\ref{simplest}, $\la \Tr X^4 \Tr X^2 \Tr X^2 \Tr X^2 \ra$,
we can see the coupling constant  ${\cal G}(4,2,2,2)$ is indeed zero
by the following argument. The field $s_2$, corresponding to $\Tr X^2$,
is in the same multiplet as the graviton (in the lowest
Kaluza-Klein level), whereas $s_4$,  corresponding to $\Tr X^4$, is in
a higher Kaluza-Klein level. On the other hand, it is believed that
Type IIB supergravity on $AdS_5\times S_5$ can be
consistently truncated to include only the multiplet of the
graviton, \ie, the field content of the gauged \N=8,
five-dimensional supergravity. Consistency requires that the
equations of motion of the untruncated theory do not contain terms
that are linear in higher Kaluza-Klein modes. Hence it implies, in
particular, that ${\cal G}(4,2,2,2)$ vanishes. More generally, the
results in \cite{West} and 
our argument for the exchange diagrams suggest that all the
next-to-extremal quartic couplings of chiral primaries are zero.

\section{Conclusion}

We have proved that next-to-extremal correlation functions of
single-trace chiral primary operators do not receive quantum
corrections at order $g^2$ in perturbation theory. This result
supports the (non-perturbative) results in~\cite{West} and
generalizes them in the sense that it applies to correlators of any
number of points.

Although we have dealt with
single-trace operators for simplicity, 
next-to-extremal correlators of multi-trace chiral primaries have an anologous 
non-renormalization property.
This can easily be shown using  the non-renormalization
theorems for two- and three-point functions of multi-trace primary
operators proven in~\cite{Skiba}, and by noting that the relevant property
for the vanishing of the diagrams that do not factorize into a
two-point and a three-point function is that each operator has to be totally
symmetric in its colour indices. Multi-trace operators have this property as well
due to the general fact that
primaries are symmetric tensors. Moreover, the $SU(4)$ structure is also
the same as for single-trace operators.

We have  studied  next-to-extremal correlation functions in
AdS supergravity as well. As in our field theory results, we have found that 
the exchange diagrams reduce to a
product of two-point functions. Note that for $I_1$ in (\ref{int1})
this occurs in a way that resembles the 
field-theoretical calculation of diagrams $a$ and
$b$ in Figs.~\ref{fig3} and~\ref{fig5} in Sections~\ref{simplest}
and~\ref{fourpoints}: the contribution (\ref{int1}) factorizes into a
two-point and a three-point function, which both have a free-field
form. On the other hand, 
$I_2$ in (\ref{int2}) corresponds to a contribution which does not
factorize into a 
two- and a three-point function, and the fact that it vanishes
independently of its detailed space-time structure seems
related to the fact that the non-factorizing field theory diagrams
discussed in Sections~\ref{simplest} and~\ref{fourpoints} are zero as
well. Of course, the AdS calculation is conjectured to describe the
strong coupling regime, 
to which perturbation theory does not apply. Nevertheless, we see that
the symmetries of the theory and the condition of next-to-extremality
enter the calculations in a similar way both at weak and at strong
coupling, such as to lead to agreeing results in both cases.

Very recently, the quartic couplings necessary for a detailed
evaluation of the  
contact diagrams within AdS
supergravity have been calculated in \cite{quartic}. 
At least in the simplest case it is easy to see 
that there is agreement with the discussion here:
The authors of \cite{quartic} find that the
quartic coupling involving three $s_2$ and a chiral primary in a
higher Kaluza-Klein level vanishes, which agrees with consistent
truncation of type IIB supergravity in $AdS_5\times S_5$. Therefore
our AdS calculation implies that the correlator $\la \Tr X^4 \Tr X^2
\Tr X^2 \Tr X^2 \ra$ 
has a free-field form for large $g$ and large $N$. On the other hand,
our perturbative result is valid for any $N$ and thus supports the
idea of consistent truncation at the quantum level.
Furthermore, according
to~\cite{West} the non-renormalization of next-to-extremal four-point
functions is a non-perturbative effect. Hence we expect all the
next-to-extremal quartic couplings in AdS supergravity to vanish, 
not just the simplest one discussed above.
It would be interesting to see whether this agrees with the results
in \cite{quartic}. This check requires making
field redefinitions as the ones carried out in \cite{quartic} for the
extremal case, and will be considered in future work. Moreover, 
our results suggest as a stronger conjecture that next-to-extremal
couplings of $n$ fields have to vanish as well. 

The consideration of instanton configurations provides further
non-perturbative evidence 
for the non-renormalization of next-to-extremal correlators. 
In~\cite{Bianchi}, an instanton calculation has been performed for
extremal correlators. 
The non-perturbative non-renormalization of
extremal correlators was checked by computing the contribution from
any instanton sector at leading order in the semiclassical
expansion. These contributions were proven to vanish using both the
systematics of gaugino zero-modes in the multi-instanton
background~\cite{instanton} and
the fact that any extremal correlator of irreducible representations
in the $SU(3)\times U(1)$ decomposition of $SU(4)$ is related by
$SU(4)$ transformations to a correlator of the form $\la z^{k_1}(x_1)
\bar{z}^{k_2}(x_2) \cdots \bar{z}^{k_n}(x_n) \ra$. In the
next-to-extremal case considered here there are
$(n-1)(n-2)/2$ invariant 
contractions of the flavour indices. All the possible invariants are
generated by the set of correlators in which the first operator is in
a maximal weight representation: $\la z^{k_1}(x_1)
(\bar{z}^{k_2-1}z)(x_2) \bar{z}^{k_3}(x_3) \cdots \bar{z}^{k_n}(x_n)
\ra$, \ldots, $\la z^{k_1}(x_1) \bar{z}^{k_2}(x_2) \cdots
\bar{z}^{k_{n-1}}(x_{n-1}) (\bar{z}^{k_n-1}z)(x_n) \ra$, where
$\bar{z}^{k_i-1}z$ denotes a traceless symmetric tensor of $SU(3)$
built out of $k_i-1$ fields $\bar{z}$ and one field $z$. In particular
we may study correlation functions for maximal weight
components in each tensor, such as $\la (z^1)^{k_1}(x_1)
((\bar{z}^1)^{k_2-2}\bar{z}z)(x_2) (\bar{z}^1)^{k_3}(x_3) \cdots
(\bar{z}^1)^{k_n}(x_n) 
\ra$. The  
multi-instanton contribution to these correlators can be proven to
vanish using an argument analogous to the one used in~\cite{Bianchi}:
In \N=4 SYM a multi-instanton configuration has sixteen exact gaugino
zero-modes $\zeta_\alpha^A$, 
corresponding to the eight
supersymmetry and eight superconformal transformations broken by the
instanton solution~\cite{zeromodes}. Solving the equations of
motion of the scalar fields in the instanton background, 
one finds (to order $g^2$) that
the scalars are bilinear in the zero-modes and that $z^1$
($\bar{z}^1$) contains 
zero-modes of flavour 0 and 1 (2 and 3) only, where 0 refers to the
spinor in the vector multiplet. On the other hand, the measure of
integration for the multi-instanton coordinates contains a factor
$d^{16}\zeta$. In order to get a non-vanishing contribution, the first
operator in the correlation function must absorb at least six of the
eight zero-modes of type 0 and 1. Hence, this operator must contain a
factor of $(\zeta)^3$. Since $\zeta$ is a two-component Grassmann
variable, $(\zeta)^m=0$ for any $m\geq 3$ and this factor (and the
entire correlator) vanishes.
This leads us to conclude that the next-to-extremal correlators do not
receive corrections from any multi-instanton sector at leading order
in the semiclassical expansion. 

All these checks support the conjecture of~\cite{West} that
next-to-extremal correlation functions of an arbitrary number of
chiral primary operators are not renormalized. 

\vspace{1cm}

{\bf Acknowledgements} \hspace{1em}

It is a pleasure to thank Dan Freedman for suggesting this problem and
for discussions. Moreover we would like to thank Leonardo Rastelli for
discussions on 
AdS supergravity.
J.E., who is a DFG Emmy Noether fellow, acknowledges funding through a
DAAD postdoctoral fellowship. M.P.V. is grateful to the MEC for a
postdoctoral fellowship.

\end{document}